\documentclass[11pt,twoside,a4paper]{article}
\usepackage[american]{babel}
\usepackage{bm}
\usepackage{amsfonts}
\usepackage{graphicx}
\usepackage{amsmath}
\usepackage{fancyhdr}
\usepackage{a4wide}
\newcommand{\nn}{\nonumber}
\newcommand{\bs}[1]{\boldsymbol{#1}}

\usepackage{color}

\newcommand{\zd}{\delta}

\newcommand{\zs}{\sigma}

\newcommand{\zg}{\gamma}

\newcommand{\zr}{\rho}
\newcommand{\za}{\alpha}

\newcommand{\zz}{\zeta}
\newcommand{\zh}{\eta}

\newcommand{\LS}[1]{ {\textbf L}^{#1} }

\newcommand{\dif}{\; \textrm d }

\newcommand{\dpp}[2]{\frac{  \partial #1 }{  \partial #2   } }
\newcommand{\dpt}[2]{\frac{  \dif #1 }{  \dif #2   } }
\begin{document}

\begin{center}
\Large{\textbf{Bose-Einstein condensation of positronium: modification of the s-wave scattering length below the critical temperature}}\\
\vskip0.5cm
	\small{O. Morandi,  P.-A. Hervieux, and G. Manfredi}	\\
\vskip0.5cm
	\textit{\textsf{Institut de Physique et Chimie des Mat\'eriaux de Strasbourg \\
23, rue du Loess,
F-67034 Strasbourg, France}}	
\vskip0.5cm
	\textit{omar.morandi@ipcms.unistra.fr}
\end{center}

\begin{center}
\begin{minipage}[h]{0.8\textwidth}
\section*{\textsf{Abstract}}
 \small
\textsf{
The production of a Bose-Einstein condensate made of positronium may be feasible in the near future. Below the condensation temperature, the positronium collision process is modified by the presence of the condensate. This makes the theoretical description of the positronium kinetics at low temperature challenging.
Based on the quasi-particle Bogoliubov theory, we describe the many-body particle-particle collision in a simple manner. We find that, in a good approximation, the full positronium-positronium interaction can be described by an effective scattering length.
%This will be used in order to quantify the corrections to the full scattering interaction in a simple manner.
Our results are general and apply to different species of bosons. The correction to the bare scattering length is expressed in terms of a single dimensionless parameter that completely characterizes the condensate.
}

\end{minipage}
\end{center}
\vspace{0.5cm}
\normalsize
%%%%%%%%%%%%%%%%   COMMENTS
%\subsection{Plan}
%
%\begin{itemize}
%  \item Bogoliubov representation: observed in 2008
%\end{itemize}
%\subsubsection{Note}
%\begin{itemize}
%  \item discuss the limit E inf in 3p interaction
%  \item  non negligible correction also to low density (because of 1/3 behav)
%  \item value n>>1 make sense. Theory may not breaks down. T goes 0. Little meaning
%  \item coeffifciente figure 1 dx linea tratteggiata approx radice di n
%\end{itemize}
%
%\begin{itemize}
%\item A good undestranding of the growth rate is crucial for the metastable Ps . Simulations swohs that teh formation of rht condesntate is below but comparable to the ps lifetime
%\item critical for BEC: condensation time. Cooling process already investigated
%\item  The Exact treatment is complex. correlation effects and BEC formation: complex framework and observable effects
%Useful some approximated theory that goes beyond to the simple bare interaction
%   \item unique feature dense BEC made by atoms
%
%\end{itemize}
%
%page
\section{Introduction}

Positronium (Ps) is the hydrogenlike bound state formed by an electron and a positron. The positronium ground state consists of two sublevels, the singlet state (para-Ps, p-Ps) and the triplet state (ortho-Ps, o-Ps), separated by a hyperfine energy gap. The particle-antiparticle structure results in the decay of the positronium via annihilation into either two or three $\zg$-ray photons. Three $\zg$-ray photons are produced by the positronium triplet spin state (lifetime in vacuum 142 ns) and two $\zg$-ray photons by the singlet spin state (lifetime in vacuum 125 ps).
Positronium has been the subject of several experimental and theoretical investigations. It finds applications of great interest in various fields such as QED \cite{Karshenboim_04}, astrophysics \cite{Guessoum _91}, the characterization of porous materials \cite{Cassidy_08,Nagashima_95},  surface composition and bulk structures \cite{Puska_94,Gorgol_12}.

One of the major challenges of today's positronium physics concerns the production of a dense gas of positronium at low temperature.
%The possibility to produce large amounts of positronium is crucial for many advanced experiments in fundamental physics.
Numerous experiments depend on the availability of a large amount of cold positronium, including gravitational measurements on antihydrogen \cite{Ferragut_10,Indelicato_14}  and the production of the molecular positronium (Ps$_2$) \cite{Cassidy_07b}.

In this paper, we focus on one of the most intriguing possibilities offered by the manipulation of dense positronium, namely the production of a Bose-Einstein condensate.
The phenomenon of Bose-Einstein condensation denotes the phase transition of a boson gas where, below the critical temperature $T_c$, a macroscopic number of bosons occupy the same quantum state: the zero momentum state for free gas or the lowest bound state for confined systems \cite{Fetter_02}.
The possibility to create a Bose-Einstein condensate made of positronium has been firstly addressed  in Ref. \cite{Platzman_94}. 
{In order to establish the feasibility of such a condensate, the process of condensate formation should be better understood. Our study helps to clarify the nature of the two body interactions during the process of the formation of the condensate. Our results indicate that due to the renormalized many-body interaction, the collisions cross section of low energy bosons increases. This result has a direct connection with the velocity at which the condensate is formed.}
Recently, the production of a 2D condensate was also investigated \cite{Mills_15}. It could be produced by implanting pulses of dense polarized positrons on the surface of a quartz crystal.  Such a 2D condensate would have properties very similar to the cold 2D polariton gas realized by Kasprzak et al. \cite{Kasprzak_06}.

The technical progress made in recent years in the production, storage and manipulation of an increasing number of positrons, suggests that the creation of a Bose-Einstein condensate of positronium atoms may be realistically possible in the near future \cite{Cassidy_10}.
{Several aspects connected to the Bose-Einstein condensation process of positronium are today under investigation. One central issue is to establish the time that is necessary to complete the condensation process. The study of the influence of the nonlinear two-body scattering interaction on the speed at which the condensate is formed has not been addressed before. For this reason, in the present work, we investigate the kinetic processes that are at the microscopic origin of the formation of the Bose-Einstein condensate.
}

%\emph{The production of a positronium gas with density of $10^{-4}$ nm$^{-3}$ is nowadays experimentally investigated ??}.
Due to the light mass of the positronium, the condensation temperature is several orders of magnitude greater than the critical temperature of the atoms usually employed in the condensation experiments.
As an example, the confinement of $10^{-4}$  Ps/nm$^{3}$ in nanometric cavities would lead to the Bose-Einstein condensation process at temperature of 30 K.
%A scenario where a BEC made of positronium could provide a source of high energetic coherent radiation has also been contemplated \cite{Loeb_86,Mills_02}.

Positronium is formed in many gases and aggregates of fine grains when an electron in an atomic or molecular orbital is captured by a positron.
A simple way to produce neutral positronium is to implant positrons into insulators, or scatter high energy positrons from metal surfaces. The incident positron beam interacts strongly with the solid and can easily capture an electron from the material and form the atom of positronium. The lifetime of an o-Ps formed in this way, is sufficiently long for it to experience many collisions with gas molecules and grain surfaces before its radiative decay. The initial positronium kinetic
energy quickly decreases via inelastic and elastic collisions \cite{Cassidy_10a,Morandi_14EJPD}.
Various theoretical and experimental studies revealed that the energy transfer from positronium to porous materials is quite efficient \cite{Cassidy_08,Nagashima_95,Nagashima_98,He_07,Zaleski_13,Mor_14PRA}. The implanted o-Ps reaches thermal equilibrium before to the annihilation time \cite{Mor_14JPB}. The thermalization time can change with the materials. In particular, in porous materials it becomes sensitive to the geometrical distribution of the cavities inside the solid \cite{Mariazzi_10}. Typical values of the thermalization time vary in the interval of 1-10 ns.

In recent years, porous silica materials have attracted much attention concerning the production of a dense positronium gas. The main reasons are that positronium thermalizes quickly in porous silica and, as a consequence of a low pick-off decay rate, o-Ps has a high survival time before annihilation, comparable with the Ps lifetime in vacuum.

In order to achieve an optimized production of positronium, new porus silica materials have been recently tested \cite{Ferragut_13,Zaleski_13} and new experimental techniques have been considered \cite{Andersen_87}. Moreover, a subject of active investigation concerns the acceleration of the positronium thermalization by the injection of remoderator gases like Ar \cite{Oka_14,Murtagh_09} or Xe \cite{Shibuya_13,Mor_15JP} at high pressure inside the porous materials.

{
In this paper, we study the modification of the hard-sphere boson-boson collision process induced by the presence of a condensate. We develop an approximation procedure where the complex highly nonlinear two-particle interaction is expressed in terms of an effective scattering length.
We describe the two-body collisions in a simple form with a clear physical interpretation. In order to proceed, we approximate the complex two-body collision integrals with some hard-sphere collision operators. We assume that the effective scattering length $a_{eff}$  varies with the quasi-particle energy. The value of   $a_{eff}$ is obtained by solving a variational problem.
The physics of the binary collisions is interpreted in a simple manner.
Depending on whether the boson energy is above or below a certain threshold, the noncondensed bosons can be classified in two groups. Above the energy threshold, the  bosons are essentially unaffected by the presence of the condensate. They collide by hard-sphere interaction and their scattering length is equal to the bare scattering length $a_0$. In the opposite limit, the low energy bosons interact with a modified scattering length. The value of the effective interaction is approximatively constant for all energies below the threshold.}

\section{Model}

\subsection{Bogoliubov-Baliaev-Popov theory}

The Bose-Einstein condensation of positronium is a kinetic process where the atoms lose energy via interaction with surfaces or impurities and decay to the ground level. Since positronium is a metastable atom it is possible to detect the evolution of the kinetic energy of the gas during the cooling process. The evolution of the positronium speed may be observed by two different techniques: The angular correlation technique \cite{Takada_00}, which measures the coincidence of two  photons produced by the positron annihilation and the Doppler-broadened spectra technique \cite{Chang_87}, which measures the energy of the emitted photons.
The  experimental detection of the gas evolution would offer a precious opportunity to investigate the validity of the kinetic theory of the condensation dynamics.

The Bose-Einstein condensate is a quantum mechanical state that extends over macroscopic distances. The detailed description of the formation of the condensate requires a fully quantum formalism.
The simple particle-particle hard-sphere interaction is usually considered as the only relevant interaction between the bosons.
However, the full quantum treatment of the boson gas dynamics indicates that the many-body quantum correlations induced by the presence of the condensate modify  the interaction between two low-energy bosons.

When the temperature of a gas of bosons decreases under a critical value $T_c$ a macroscopic number of bosons occupy the lowest energy level and form a Bose-Einstein condensate. At finite temperature, the condensate coexists with a gas of bosons whose thermal energy dispersion follows the Bose-Einstein distribution with zero chemical potential (we will denote this ensemble of bosons as ``noncondensed gas").
The relevant two-particle interaction of most boson systems that have been considered for the experimental production of a Bose-Einstein condensate (with the exception of bosons that are complex excitations of solid state systems like, for example, the polariton), is the hard sphere  s-wave collision. Hereafter, we will denote by $a_0$ the bare scattering length.

In this work, we focus on the modification of the boson-boson bare interaction below the condensation temperature.

According to the Bogoliubov theory, the quantum mechanical correlation with the condensate modifies significantly the microscopic boson-boson interaction inside the noncondensed gas.
A new type of interaction,  hereafter denoted as ``condensed-noncondensed (NC) interaction", becomes relevant (for a general introduction to this subject see   \cite{Griffin_book}). From a physical point of view, the NC interaction takes into account the microscopic processes whereby a boson is exchanged between the condensate and the noncondensed gas.
Despite the fact that the total number of bosons is conserved, the scattering NC process is formally described by a two-particle collision process with creation (which physically corresponds to the ejection of a boson from the condensate into the gas) or annihilation of a boson (which indicates the capture of one free boson by the condensate). Mathematically, it is described by a three-density Boltzmann collision integral containing some modifications of the  scattering coefficients and the density of states.
%

%In the framework of the many-body theory, the bosons in the noncondensed gas are represented (in terms of / by) quasi-particle excitations.
The many-body theory of a gas of bosons interacting with a condensate was initially developed by Bogoliubov, Beliaev and Popov \cite{Bogoliubov_47,Baliev_58,Popov_87}. According to the Bogoliubov-Beliaev-Popov (BBP) theory, the  bosons in the noncondensed gas  are represented by dressed quasi-particles. Consequently, the two-particle collision takes a more complex form that differs substantially from the simple hard-sphere interaction.
Intuitively, the main corrections to the bare boson-boson interaction are expected to take place at low energy. Indeed, only the interaction between bosons whose energy is close to zero (in the Bogoliubov approach the zero of the energy is taken equal to the condensate mean field energy, so that the energy of the condensate is zero by definition) should be modified by the presence of the condensate. At high energy, the bosons become free and the main scattering interaction reduces to the bare hard-sphere interaction.

A quasi-particle with momentum $\mathbf{p}$ is a quantum-mechanical state made of the superposition of a pair of particle-hole states with momentum $\mathbf{p}$. The squared modulus of the projection of the quasi-particle state in the particle space is given by
$  \mu =  \frac{1}{2}+ \frac{1}{2\sqrt{1+\left(\frac{E_0}{E}\right)^2}}$ \cite{Fetter_02}. Here, $E$ is the quasi-particle energy, $E_0= g n_c$ where $n_c$ is the condensate density, $g$ the interaction strength given by  $g=\frac{4\pi  \hbar^2}{m} a_0$ and  $m$ is the boson mass. At low energy ($E\ll E_0$), $\mu \rightarrow $ 0.5. The quasi-particle is an equal mixture of particle and hole states. In the opposite limit, at the free-particle regime $E\gg E_0$, $\mu \rightarrow 1 $ and the quasi-particle reduces to the usual bare boson. In the same way, dressed interactions are expected to degenerate into the bare s-wave scattering for $E \gg E_0 $.

The BBP theory thus indicates the existence of a certain value of the quasi-particle energy $E_0$ under which the modification of the bare two-particle interaction becomes important. This consideration plays a crucial role for the definition of an effective scattering length.

\subsection{Kinetic theory of the positronium gas}

We discuss now the evolution equation of the quasi-particles. Since the drift motion of the quasi-particles is not relevant for our discussion,   we will assume that the gas of bosons and the condensate are uniform in space. From a mathematical point of view, the evolution of the   quasi-particle distribution function $f$ of the  noncondensed interacting bosons is described by the Boltzmann equation \cite{Kirkpatrick_85,Jackson_02,Mor_13PRA} :
\begin{align}
\dpp{f}{t} &= \mathcal{Q}  [\textsf{T},f] + \mathcal{W} [\textsf{S},f]\;. \label{ini Bz eq}
\end{align}
The first term on the right-hand side describes the collision between two bosons that belong to the noncondensed gas (NN interactions). We have:
\begin{align}
 \mathcal{Q} [\textsf{T},f]( \mathbf{p}_1)  = &  \zg \int \textsf{T} \left[\left(1+f_1 \right) \left(1+ f_2\right)  f_3  f_4  -f_1  f_2 \left(1+f_3\right) \left(1+  f_4\right) \right]   \nn  \\
 & \hspace{2cm}\times \zd\left(E_1 + E_2- E_3 - E_4 \right) \zd \left(\mathbf{p}_1 +\mathbf{p}_2-\mathbf{p}_3-\mathbf{p}_4\right) \dif\mathbf{p}_2 \dif\mathbf{p}_3 \dif\mathbf{p}_4    \;,   \label{Q}
\end{align}
where $\zg=\frac{8 a_0^2}{(2\pi)^3\hbar^3 m^2 }$, $\mathbf{p}_i$ denotes quasi-particle momentum and $\textsf{T}$ is the scattering amplitude, whose expression is given in the Appendix (Eq. \eqref{T}). We introduced the shorthand notations $f_i\equiv f (\mathbf{p}_i)$ and $E_i\equiv \mathcal{E} (\mathbf{p}_i)$, where $\mathcal{E} (\mathbf{p}_i)$ denotes the Bogoliubov quasi-particle energy $  \mathcal{E} (\mathbf{p}) =  \sqrt{\left(\frac{\mathbf{p}^2}{2m}+E_0\right)^2 - E_0^2 }$.

The second term on the right-hand side of Eq. \eqref{ini Bz eq} describes the processes where one boson is exchanged between the condensate and the non-condensate gas (NC interactions). Such processes are crucial for the description of the condensate growth. We have
\begin{align}
 \mathcal{W} [\textsf{S},f] ( \mathbf{p}_1)  =&  2\xi \int    \zd\left(E_1 + E_2 - E_3  \right)   \zd \left(\mathbf{p}_1 +\mathbf{p}_2-\mathbf{p}_3 \right)  \left[ \left(1+f_1 \right) \left(1+ f_2\right)  f_3 - f_1  f_2 \left(1+f_3\right)  \right] \textsf{S}  \dif\mathbf{p}_2 \dif\mathbf{p}_3    \nn  \\
%%%
&  +\xi  \int   \zd\left( E_1 - E_2 - E_3 \right) \zd \left(\mathbf{p}_1 -\mathbf{p}_2 -\mathbf{p}_3\right) \left[  \left(1+f_1 \right)  f_2  f_3 - f_1  \left(1+f_2\right) \left(1+f_3\right)  \right]\textsf{S}  \dif\mathbf{p}_2 \dif\mathbf{p}_3  \; , \label{W}      %
\end{align}
where $\xi = \frac{8 a_0^2 n_c}{ m^2 }$ and $\textsf{S}$ is the correspondent scattering amplitude, whose expression is given in Eq. \eqref{S} in the Appendix.
In many relevant cases, the distribution function of the quasi-particles is isotropic with respect to the momentum. Consequently, in Eq. \eqref{ini Bz eq}  it is more useful to use the quasi-particle energy $E_i = \mathcal{E} (\mathbf{p}_i)$
instead of the momentum. In our discussion, we always assume that the bosons are at equilibrium so that the isotropic assumption applies. In this case, the expressions of the Boltzmann collision kernels $\mathcal{Q}  [\textsf{T},f]$ and $\mathcal{W} [\textsf{S},f]$ simplify considerably. They are given in the Appendix.

The derivation of the Boltzmann integrals $ \mathcal{Q} [\textsf{T},f] $ and $ \mathcal{W} [\textsf{S},f] $ is quite complex. In particular, the derivation of the scattering coefficients $\textsf{S}$ and  $\textsf{T}$ requires cumbersome calculations \cite{Mor_13PRA,Griffin_book,Imamovic_01}.

{Such scattering kernels are obtained in a  quantum many-body framework. The condensate comprises  a macroscopic number of atoms. However,  the condensate wave function is not an eigenfunction of the number operator. For this reason, the number of atoms ins the condensate is not a well-defined quantity. The fluctuations of the number of atoms contained in the condensate lead to an entanglement between  the atoms inside and outside the condensate. This phenomenon is  at the origin of the highly nonlinear two-body interactions described by the scattering coefficients $\textsf{S}$ and  $\textsf{T}$. }
In the standard description of the two-body collision processes, the condensate-noncondensate collision integral $ \mathcal{W} [\textsf{S},f] $ is not present. The condensate-noncondensate collision integral $ \mathcal{W} [\textsf{S},f] $ has no analogous term in the standard two-body collision processes. It is a direct consequence of the Bogoliubov quasi-particle transformation and is an original result of the BBP theory.
However, Eq. \eqref{W}  can be justified as follows. We assume that the distribution function that describes both the condensate and the gas of noncondensed bosons can be written as \cite{Banyai_00}
\begin{align}
f(\mathbf{p})     =& f'(\mathbf{p})  + \zd(\mathbf{p})  (2\pi\hbar)^3 n_c \; , \label{ansatz f}
\end{align}
where $f'$ is an integrable and regular function except maybe in $\mathbf{p}=0$. The function $f'$ describes the noncondensed gas and the term with the Dirac delta the condensate. We use the ansatz of Eq. \eqref{ansatz f} in Eq. \eqref{Q}. By formally developing the result up to the fist order in the condensate density $n_c$, we obtain the operator $ \mathcal{W} $ except for the term $\textsf{S} $, which cannot be derived in such an elementary manner.

The main difficulty related to the application of Eq. \eqref{ini Bz eq} to some real situation arises from the presence of the collision scattering kernels $ \textsf{T}$ and $ \textsf{S}$ in the Boltzmann collision integrals. They characterize the microscopic boson-boson collision and are nontrivial functions of the energy of the quasi-particles before and after the collision.
It would be convenient to have a simple approximation of such scattering integrals. The main result of our analysis is  to show that to a good approximation, it is possible to replace the complex two-body interaction by a hard-sphere interaction.

%
%
%The appropriate scaling of the system is easily found by observing that dependence on the momentum of the quasi-particles of the scattering coefficients is provided by the product $\frac{g n_c}{E}$, where $E=\mathcal{E}(p)$ is the energy of the boson. This functional is a general consequence of the BPB perturbation theory. According to that, the corrections to the bare b-b interaction at all order of the interaction strength are expressed in terms of the projection of the Bv creation operator on the bare particle-hole basis $U^2=    \frac{1}{2}\sqrt{ 1 + \left(\frac{g n_c}{E} \right)^2 }+ \frac{1}{2}$.
%\begin{itemize}
%  \item In order to illustrate we consider a (maybe) anomalous Feynmann diagram: the dressed bv interaction strength is proportional to U
%\end{itemize}
%In particular, the coefficients $ \textsf{T} (U(E_1),U(E_2),U(E_3),U(E_4))$ and $  \textsf{S}(U(E_1),U(E_2),U(E_3))$ are polynomials of the coefficient $U$ for respectively three and four qb energies.

In order to proceed, it is useful to rewrite  Eq. \eqref{ini Bz eq} in term of dimensionless variables.
As already mentioned, $E_0 = g n_c$ is the typical value of the  energy where the modifications of the boson-boson collision take place. This is confirmed by the fact that the scattering coefficients $\textsf{T}$ and $\textsf{S}$ depend on the quasi-particle energy via the product $\frac{g n_c}{E}$ (explicitly $  \textsf{T} \left(\frac{E_0}{E_1},\frac{E_0}{E_2},\frac{ E_0}{E_3},\frac{E_0}{E_4} \right)$ and $ \textsf{S} \left(\frac{E_0}{E_1},\frac{E_0}{E_2},\frac{E_0}{E_3} \right)$).
This consideration suggests the use of the following scaled variables for the energy $E'=E / k_B T $, the momentum $p'=p /\sqrt{k_B T}$, and the condensate density $\overline{n}=\frac{g n_c}{k_B T} $. Here, $T$ denotes the temperature. In the scaled variables the equation becomes (for simplicity, we drop the prime in our notation)
\begin{align}
 \mathcal{Q} [\textsf{T},f]( \mathbf{p}_1)  = &   \zg\left(  k_B T\right)^2 \int  \textsf{T} \left(\frac{ \overline{n}}{E_1},\frac{\overline{n}}{E_2},\frac{\overline{n}}{E_3},\frac{\overline{n}}{E_4} \right) \left[\left(1+f_1 \right) \left(1+ f_2\right)  f_3  f_4  -f_1  f_2 \left(1+f_3\right) \left(1+  f_4\right) \right]   \nn  \\
 & \hspace{2cm}\times \zd\left(E_1 + E_2- E_3 - E_4 \right) \zd \left(\mathbf{p}_1 +\mathbf{p}_2-\mathbf{p}_3-\mathbf{p}_4\right) \dif\mathbf{p}_2 \dif\mathbf{p}_3 \dif\mathbf{p}_4    \;,   \label{Qren}
\end{align}
and similarly for the term  $\mathcal{W} [\textsf{S},f]$ with the substitution $\xi \rightarrow \frac{8 a_0 \overline{n}}{ m 4\pi \hbar^2 } $. In our analysis, the scaled condensate density  $\overline{n}=\frac{g n_c}{k_B T} $ plays a relevant role. It contains the physical parameters that characterize the condensate (density $n_c$, temperature $T$ and interaction strength $g$) and will be used in order to relate our results to the experimental conditions.

Now, we write $\overline{n}$ in a form that is more suitable for the physical interpretation. The following formula $T_c=\frac{2\pi \hbar^2}{mk_B} \left(\frac{n}{2.612}\right)^{2/3}$ relates the condensation temperature to the total density of the bosons $n$. Moreover, the ratio between the density of condensed bosons $n_c$ and the total density $n$ is given by  $\frac{n_c}{n}=1-\left(\frac{T}{T_c}\right)^{3/2}$ (see, e.g., \cite{Fetter_02}). Simple manipulations lead to
\begin{align*}
\overline{n}=& 3.79 \left( \frac{T_c}{T} -  \sqrt{\frac{T}{T_c}} \right) \; a_0 \;  n^{1/3}\; ,
\end{align*}
where we used $g=4\pi\hbar^2a_0/m$. Of particular interest is the case where the condensate density is of the same order of the total density of the gas. For instance, when the system is composed by a $50/50$ density mixture of condensed and noncondensed bosons ($\frac{n_c}{n}=0.5$), the previous formula becomes
\begin{align*}
\overline{n}=& 3.01  \; a_0 \;  n^{1/3}
\end{align*}
It is worth noting that $\overline{n}$ is directly related to the relevant dimensionless parameter $n_c a_0^3$ which is usually considered for the study of the many-body expansion of the two-particle interaction \cite{Fetter_02}.
{Our results are based on the dressed scattering interaction obtained from the BBP theory. The validity of the BBP theory is restricted to the case of diluted gases for which $n a_0^3\ll 1$. Accordingly, our results are limited to small values of the scaled condensed density $\overline{n}$. }

%1\notaf{Important the 1/3 factor, makes the corrections (more persistent and relevant also to small) less sensible to small density (usually work conditions)}
Typical values of $\overline{n}$ for several atomic systems are indicated in Tab. \ref{tab n}.
\begin{table}[h]
\begin{center}
 \begin{tabular}{l|l|l|l}
      \hline
 Atom & Density (nm$^{-3}$) & Scattering length (nm)  & $\overline{n}$   \\
 o-Ps  & $10^{-7}-10^{-3}$  & 0.16 \cite{Ivanov_02}&   $  10^{-3}-4 \times10^{-2} $\\
 $^{87}$Rb &$10^{-9}$ &5.5 \cite{Pethick_book} & $1.7\times10^{-2}$\\
 $^{23}$Na &$10^{-9}$ &4.5 \cite{Pethick_book} & $1.4\times10^{-2}$ \\
      \hline
\end{tabular}
\caption{\label{tab n} Typical values of the scaled density $\overline{n}$ in some realistic cases.}
\end{center}
\end{table}

\section{Minimization technique}

In this section, we describe an optimization technique that is useful to simplify the two-boson collision process.
In order to introduce our approach, it is useful to consider a textbook result. Let us consider the classical Boltzmann equation for a rarefied gas subject to some elastic collisions described by the transition rate $\textsf{W}(\mathbf{p}_1,\mathbf{p}_2)$, where $\mathbf{p}_1$ and $ \mathbf{p}_2$ are respectively the pre- and post- collision momenta. The master equation for the boson density is
\begin{align}
\dpp{f}{t}  &=   \int \textsf{W}(\mathbf{p}_2,\mathbf{p}_1) f( \mathbf{p}_2)\dif\mathbf{p}_2   - \frac{f(\mathbf{p}_1 ) }{\tau(\mathbf{p}_1)}\label{BE ex}
\end{align}
where
\begin{align}
\tau^{-1} (\mathbf{p}_1)  &= \int \textsf{W}(\mathbf{p}_1,\mathbf{p}_2)    \dif\mathbf{p}_2 \;. \label{tau ex}
\end{align}
The last term of Eq. \eqref{BE ex} is the so-called ``loss term", which takes the form of the Bhatnagar-Gross-Krook (BGK) relaxation time. Such a simple form allows to interpret $\tau^{-1}$ as the collision frequency of a particle with momentum $\mathbf{p}_1$.  This the natural way to proceed in the case of the linear Boltzmann equation \eqref{BE ex} and makes the $\tau^{-1}$ term simpler to analyze than the full collision coefficient $\textsf{W}(\mathbf{p}_1,\mathbf{p}_2)$.

In the following, in analogy with Eq. \eqref{tau ex}, we replace the coefficients $\textsf{S}$ and $\textsf{T}$ by a simpler parameter that is interpreted as an effective scattering length for the dressed quasi-particles.

We will focus only on Eq. \eqref{Q}, but the same considerations apply to Eq. \eqref{W} with obvious modifications.
$ \mathcal{Q} [\textsf{T},f] ( E_1)$ is a nonlinear operator. It maps the pair $(\textsf{T},f)$ to the function $h= \mathcal{Q} [\textsf{T},f]$. Our strategy is to replace the four-energy function $\textsf{T}(E_1 ,E_2 ,E_3 ,E_4 )$ with a single function $\zs(E_1 )$. The function $\zs$ is chosen in such a way that the distance between the new function $h' \equiv \mathcal{Q} [\zs,f]$ and $h$ is minimum. The distance is calculated with the following $L^2$ norm
\begin{align}
  \| h \|_{\LS{2}}  &\equiv \sqrt{\int h^2(E)  \zr^2 (E)  \dif E }\;.
\end{align}
Here, $\zr$ denotes the density of states. The minima should  be taken for all the functions $f$ that belong to a certain functional space. For our purposes, it is more convenient to chose $f$ on the basis of some physical considerations. {The Bose-Einstein condensate is characterized by the existence of macroscopically occupied  quantum states and, strictly speaking, does not rely on thermal equilibrium. However, the Bose-Einstein condensate of weakly interacting gases is typically formed at the thermal equilibrium.} In these cases the distribution function of noncondensed bosons is the Bose-Einstein function $f_{BE}$ with zero chemical potential (in our normalized units $f_{BE}=[e^E-1]^{-1}$). This would suggest to evaluate the minimum of the norm by setting $f=f_{BE}$. However, by definition, the Bose-Einstein distribution belongs to the kernel of the collision operator irrespectively of the choice of the scattering parameter. In this case, our minimization procedure would lead to the trivial solution $\zs=0$. This problem is solved if, in analogy with Eq. \eqref{BE ex}, we evaluate the minima only for the loss term (out-scattering) of the collision integral in Eq. \eqref{Q} (the term with the minus sign). We remark that, since at equilibrium the two out- and in- scattering terms balance each other, the same calculation applied to the gain term would not change the final result. Hereafter, we will denote with the superscript $l$ the integral collision kernel containing only the loss terms.

In the end, our approach leads to the following optimization problem: to find $\zs(E)$ that minimizes the norm $   \| \mathcal{Q}^{l} [\textsf{T},f_{BE}]-\mathcal{Q}^{l} [\zs,f_{BE}]   \|_{\LS{2}} $. The variational calculation gives
%\begin{align*}
%   \| \left(\mathcal{Q} [\textsf{T}_{nn},f]-\mathcal{Q} [\zs,f]\right) \zr  \|_{\LS{2}}^2 =& \int_0^\infty  \left(\mathcal{W} [\textsf{T}_{nn},f]-\mathcal{W} [\zs,f]\right)^2 \zr^2(E_1) \dif E_1 \\ \equiv &\int_0^\infty \left[ \int_0^\infty  \int_0^{E_1+E_2} \left(\textsf{T}_{nn}  -\zh \right) F(E_1,E_2,E_3)\zr (E_1) \dif E_2\dif E_3  \right]^2 \dif E_1
%\end{align*}
\begin{align*}
  \delta  \|  \mathcal{Q}^{l} [\textsf{T},f_{BE}]-\mathcal{Q}^{l} [\zs,f_{BE}]   \|_{\LS{2}}^2 =&-2\int_0^\infty  \left( \mathcal{Q}^l [\textsf{T}-\zs,f_{BE}]    \right) \left( \mathcal{Q}^l[\zd \zs,f_{BE}]   \right) \zr^2(E_1) \dif E_1 \; .
\end{align*}
The minimum is thus given by
\begin{align*}
 \frac{\delta}{\delta \zs} \left\| \mathcal{Q}^l [\textsf{T},f_{BE}]-\mathcal{Q}^l [\zs,f_{BE}] \right\|_{\LS{2}}=0 \; .
\end{align*}
We obtain
\begin{align}
\zs(E) = \frac{\mathcal{Q}^l [\textsf{T},f_{BE}] }{\mathcal{Q}^l [1,f_{BE}]} \; .\label{zs gen for}
\end{align}
It is useful to note that this result agrees with Eq. \eqref{tau ex}. Equation \eqref{zs gen for} applied to  the  rarefied Boltzmann gas \eqref{BE ex} provides $\zs=\tau^{-1}$.

In our procedure, the optimum function $\zs$ is a local minimum with respect to the energy. A stronger requirement would be to find a constant $\zs_0$ such that the problem has a global minimum. This can be easily obtained by replacing  $\zs$ with a constant and the operator $\frac{\delta }{\delta \zs } $ with the ordinary derivative $\dpt{}{\zs}$. In this case, the results is
\begin{align*}
{\zs}_0 = \frac{\int_0^\infty\mathcal{Q}^l [\textsf{T},f_{BE}] \mathcal{Q}^l [1,f_{BE}]  \zr^2(E_1) \dif E_1}{\int_0^\infty \left(\mathcal{Q}^l [1,f_{BE}]  \zr (E_1)  \right)^2 \dif E_1}\; .
\end{align*}
However, since some interesting physical insight emerges by the analysis of the behavior of $\zs$ as a function of the quasi-particle energy, in the following we will make use only of Eq. \eqref{zs gen for}.

\section{Results }
\subsection{NN scattering}
\begin{figure}[h]
\begin{center}
\includegraphics[height=0.35\columnwidth,width=0.49\columnwidth]{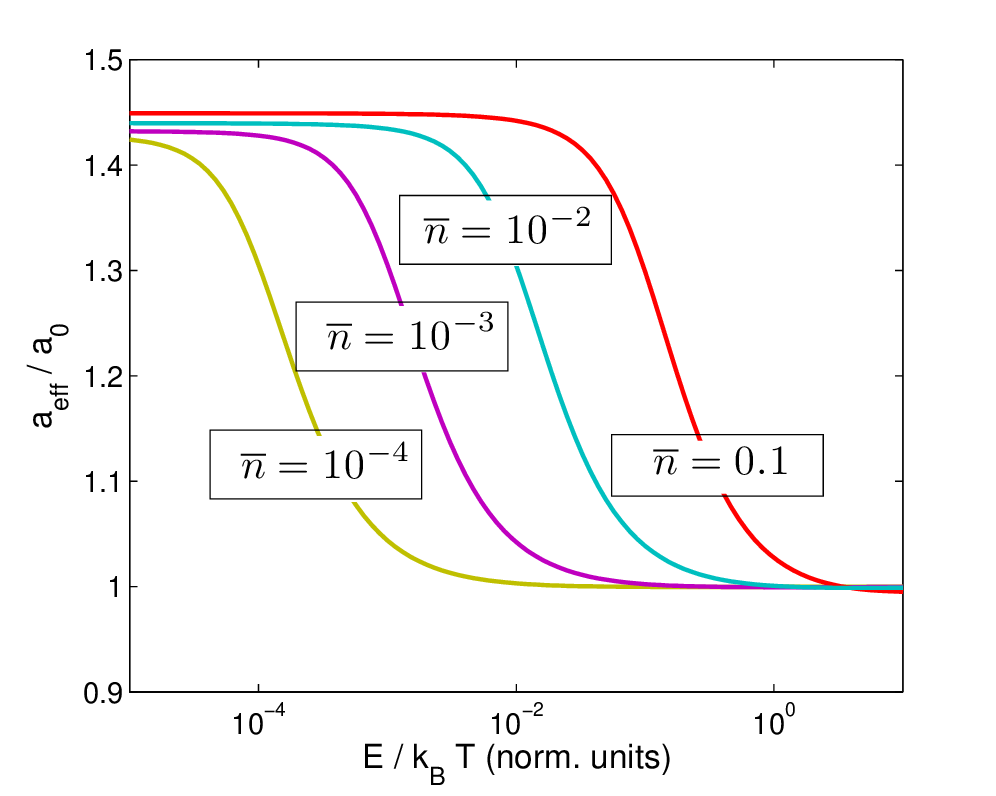}
%\notaf{sim in file code\4p_zs_E \ sim_par  }
\includegraphics[height=0.35\columnwidth,width=0.49\columnwidth]{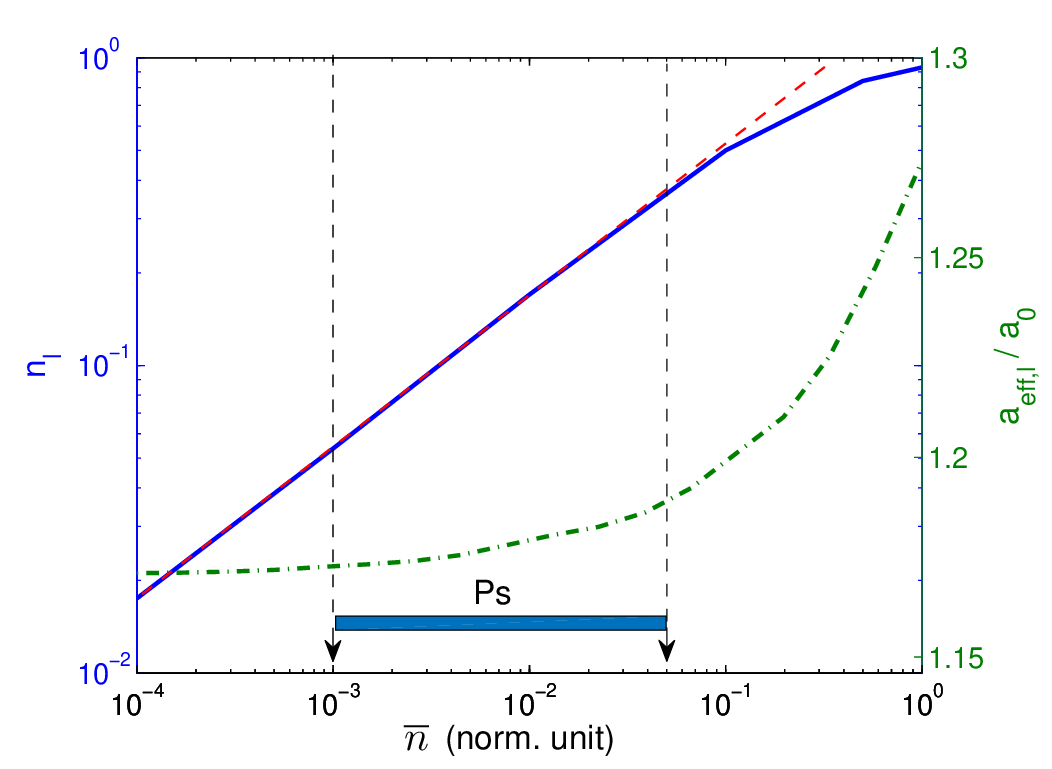}
%\notaf{sim in file code\4p_zs_E \ sigm_medio  }
\caption{\label{fig zs_E} Left panel: Effective scattering length $\za_{\textsf{T} }$ as a function of the energy, for different normalized  condensate densities $\overline{n}$. Right panel: Fraction of bosons with modified scattering length (blue continuous curve, left vertical axis), together with the approximation  $n_l \propto \sqrt{\overline{n}}$ (red dashed line); the green dot-dashed curve represents the effective scattering length (right vertical axis).}
\end{center}
\end{figure}
We apply our minimization procedure to the NN collision kernel. Equation \eqref{Q} shows that the collision integral is proportional to the square of the scattering length $a_0$. It is convenient to define $\zs(E) \equiv  \za_{\textsf{T} }^2$ and interpret the quantity  $a_{eff}=a_0  \za_{\textsf{T} }$ as an effective scattering length. The full collision dynamics is thus approximated by a hard-sphere interaction. The only modification is the introduction of an energy dependent effective scattering length. We have
\begin{align}
   \za_{\textsf{T} }^2 (E_1) =&  \frac{    \int  \textsf{T} \left(\frac{ \overline{n}}{E_1},\frac{\overline{n}}{E_2},\frac{\overline{n}}{E_3},\frac{\overline{n}}{E_4} \right) \mathcal{M} \dif\mathbf{p}_2 \dif\mathbf{p}_3 \dif\mathbf{p}_4  }{  \int \mathcal{M} \dif\mathbf{p}_2 \dif\mathbf{p}_3 \dif\mathbf{p}_4  }\;,  \label{4p zs_E}
\end{align}
where
\begin{align*}
  \mathcal{M}  =& \frac{ \zd\left(E_1 + E_2- E_3 - E_4 \right) \zd \left(\mathbf{p}_1 +\mathbf{p}_2-\mathbf{p}_3-\mathbf{p}_4\right)}{ \left(e^{  E_2}-1 \right)     \left(1-e^{- E_3} \right)      \left(1-e^{- E_4} \right) }\; .
\end{align*}
After some algebra [we use Eq. \eqref{Q simp} in the Appendix], Eq. \eqref{4p zs_E} simplifies to:
\begin{align}
   \za_{\textsf{T} }^2  (E_1) =&  \frac{   \int_0^\infty  \int_0^{E_1+E_2}   \textsf{T}   \left(\frac{ \overline{n}}{E_1},\frac{\overline{n}}{E_2},\frac{\overline{n}}{E_3},\frac{\overline{n}}{E_3,E_1+E_2-E_3} \right)  \mathcal{M}'   \dif E_2\dif E_3   }{  \int_0^\infty   \int_0^{E_1+E_2}\mathcal{M}'  \dif E_2\dif E_3   } \;,\label{4p zs_E sim}
\end{align}
where
\begin{align}
   \mathcal{M}'  =& \frac{   \zz (E_1,E_2,E_3,E_1+E_2-E_3)}{ \left(e^{  E_2}-1 \right)     \left(1-e^{- E_3} \right)      \left(1-e^{- E_4} \right) }
\end{align}
and $ \zz$ is given by Eq. \eqref{zz}.
The result of the calculation is shown in Fig. \ref{fig zs_E}. We depict $\za_{\textsf{T} }$ as a function of the scaled quasi-particle energy (note that we use a logarithmic scale for the energy). Each curve is related to a different value of the normalized condensate density $\overline{n}$.  The effective scattering length shows a monotonic two-step behavior. As expected, at high energy the quasi-particle loses its character of mixed quantum state and reduces to the simple bare boson. Accordingly, $\za_{\textsf{T} }$ goes to unity and the effective scattering length reduces to $a_0$.  As already pointed out, the value $\overline{n}$ discriminates between low and high quasi-particle energy. When the energy decreases, around $E=\overline{n}$, the function $\za_{\textsf{T} }$ increases rapidly and saturates to a value around $\sqrt{2}$. From Eq. \eqref{4p zs_E sim} it easy to see that for $E\ll\overline{n}$ and $\overline{n}$ going to zero, $\za_{\textsf{T} }\rightarrow \sqrt{2}$.
The relevant limit for low energy scattering is thus
\begin{align*}
\lim_{\overline{n} \rightarrow 0}   \lim_{E_1\rightarrow 0}  \za_{\textsf{T}}^2 =&    \lim_{\overline{n} \rightarrow 0} \lim_{E_1\rightarrow 0} \textsf{T}  = 2\;.
\end{align*}
This two-step behavior suggests the following interpretation of the results. We divide the noncondensed bosons into two populations characterized by two different scattering lengths. Low energy ($E<\overline{n}$) bosons have an augmented scattering length of around $\sqrt2$, while high energy ($E>\overline{n}$) bosons keep the bare interaction $a_0$. In order to evaluate the ratio between the density of low and high energy bosons, we calculate the fraction $n_l$ of bosons whose effective scattering length is more than 5 $\%$ higher than the bare interaction.
\begin{align}
  n_l= \frac{1}{n}\int_{\za_{\textsf{T}}>1.05} f_{BE}(\mathbf{p}) \dif \mathbf{p} = \frac{1}{n}\int_{\za_{\textsf{T}}>1.05} \frac{1}{e^E-1}  \zr(E) \dif E \;.
\end{align}
Here, $n$ is the density of noncondensed bosons. The mean scattering length seen by the low energy population is
\begin{align}
  a_{eff,l}= \frac{a_0}{n_l}\int_{\za_{\textsf{T}}>1.05}\za_{\textsf{T}}  f_{BE}(\mathbf{p}) \dif \mathbf{p}  = \frac{a_0}{n_l}\int_{\za_{\textsf{T}}>1.05}\za_{\textsf{T}} (E) \frac{1}{e^E-1}  \zr(E) \dif E \;. \label{za eff l}
\end{align}
The result of the calculation is shown in Fig. \ref{fig zs_E} (right panel). The interval of values that are relevant  to the positronium condensation process are indicated by vertical lines. However, our result is quite general and can be applied to different boson systems.  The only parameter that describes the condensate is $\overline{n}$. In order to discuss our results, let us consider a Bose-Einstein condensate made of ortho-positronium with density of $10^{-3}$ nm$^{-3}$. According to Table \ref{tab n}, $\overline{n}\simeq 4\times 10^{-2}$. From Fig. \ref{fig zs_E} (right panel) we see that around 30 $\%$ of the positronium atoms (continuous blue curve) scatter with a scattering length of around $1.18  \;a_0 \simeq 0.2$ nm. For the remaining 70 $\%$ of the positronium atoms the scattering length is not modified. This example illustrates that, by using our approach, the corrections to the collision dynamics may be quantified in a simple way.

Our discussion is based on the saturation of $\za_{\textsf{T}} $ around $E=\overline{n}$, which can be understood by simple considerations. According to the Bogoliubov theory, the density of states of the quasi-particles is
%\begin{align}
%  \zr(E)= \frac{2\pi m \sqrt{2m}}{\sqrt{1+\left(\frac{\overline{n}}{E} \right)^2}}  \sqrt{\sqrt{E^2 +  (\overline{n})^2}  -   \overline{n}}
%\end{align} $  \mathcal{E}^{-1} (E) =   \sqrt{\sqrt{E^2 +  (\overline{n})^2}  -   \overline{n}} $
%\begin{align*}
%  \zr(E)=2\pi m \sqrt{2mE } \sqrt{\frac{\sqrt{\left(\frac{\overline{n}}{E} \right)^2 + 1}  -   \frac{\overline{n}}{E}}{1+\left(\frac{\overline{n}}{E} \right)^2} }=
%  %
%  2\pi m \sqrt{2m }\frac{E}{\sqrt{\overline{n}}} \sqrt{\frac{\sqrt{\left(\frac{E}{\overline{n}} \right)^2 + 1}  -   \frac{E}{\overline{n}}}{1+\left(\frac{E}{\overline{n}} \right)^2} }
%\end{align*}
\begin{align*}
  \zr(E)=
 \frac{\sqrt{2 } }{(2\pi)^2 \hbar^3} V m^{3/2} \frac{E}{\sqrt{\overline{n}}} \sqrt{\frac{\sqrt{\left(\frac{E}{\overline{n}} \right)^2 + 1}  -   \frac{E}{\overline{n}}}{1+\left(\frac{E}{\overline{n}} \right)^2} } \;,
\end{align*}
%\begin{align*}
%  \zr(E)=\frac{V}{(2\pi \hbar)^3}2\pi m \sqrt{2mE } \sqrt{\frac{\sqrt{\left(\frac{\overline{n}}{E} \right)^2 + 1}  -   \frac{\overline{n}}{E}}{1+\left(\frac{\overline{n}}{E} \right)^2} }=
%  %
% \frac{V}{(2\pi \hbar)^3} 2\pi m \sqrt{2m }\frac{E}{\sqrt{\overline{n}}} \sqrt{\frac{\sqrt{\left(\frac{E}{\overline{n}} \right)^2 + 1}  -   \frac{E}{\overline{n}}}{1+\left(\frac{E}{\overline{n}} \right)^2} }
%\end{align*}
where $V$ is the volume of the system. %At the equilibrium, the bosons distribution function is  $f_{BE}=1/(e^E-1)$.
We assume here for simplicity $\overline{n}<1$.  At high energy  $E\gg 1 > \overline{n}$ (corresponding to $E\gg k_B T$ for the non-scaled variables), the product $f \zr$ is dominated by the exponential decreasing. In the opposite limit  $E\ll \overline{n}$, the product $\zr f_{BE} $ goes as ${\overline{n}}^{\,-1/2}$. It is interesting to note that the function $\zr f_{BE}$ is regular in the right neighbourhood of zero. This contrasts with the standard theory of bare bosons for which the product between the Bose-Einstein distribution and the density of states goes as $E^{\,-1/2}$ around $E=0$. The real condensates that are produced in the experiments are characterized by small parameters $\overline{n}$. In this case, we can estimate the number of low energy bosons as proportional to  $\frac{1}{\sqrt{\overline{n}}} \overline{n} = \sqrt{\overline{n}}$. In order to validate our estimation, in Fig. \ref{fig zs_E} (right panel) the red dashed line depicts the curve $\sqrt{\overline{n}}$.
%\notaf{il coeff dovrebbe esssere $1/\int \sqrt{E} FB$}
Our simple estimation agrees very well with the numerical results.

\subsection{NC scattering}
\begin{figure}[h]
\begin{center}
\includegraphics[height=0.35\columnwidth, width=0.49\columnwidth]{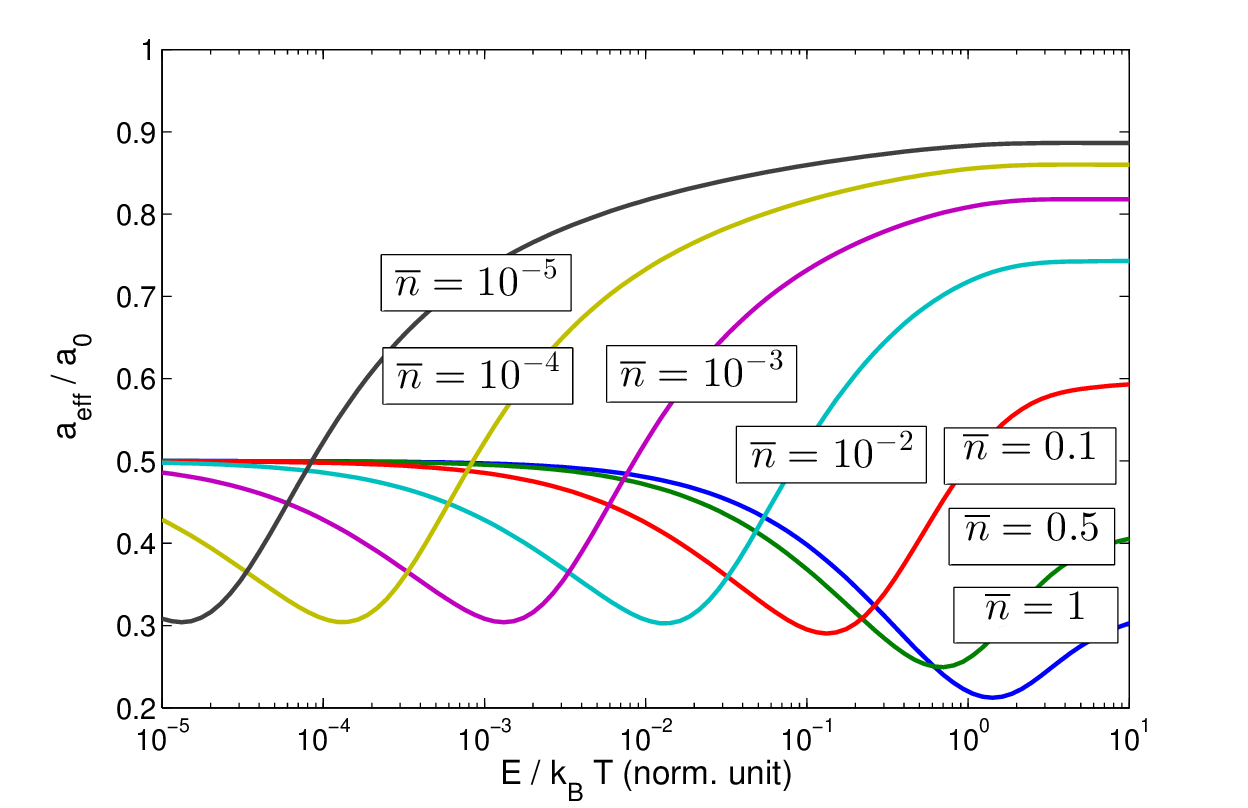}
\includegraphics[height=0.35\columnwidth,width=0.49\columnwidth]{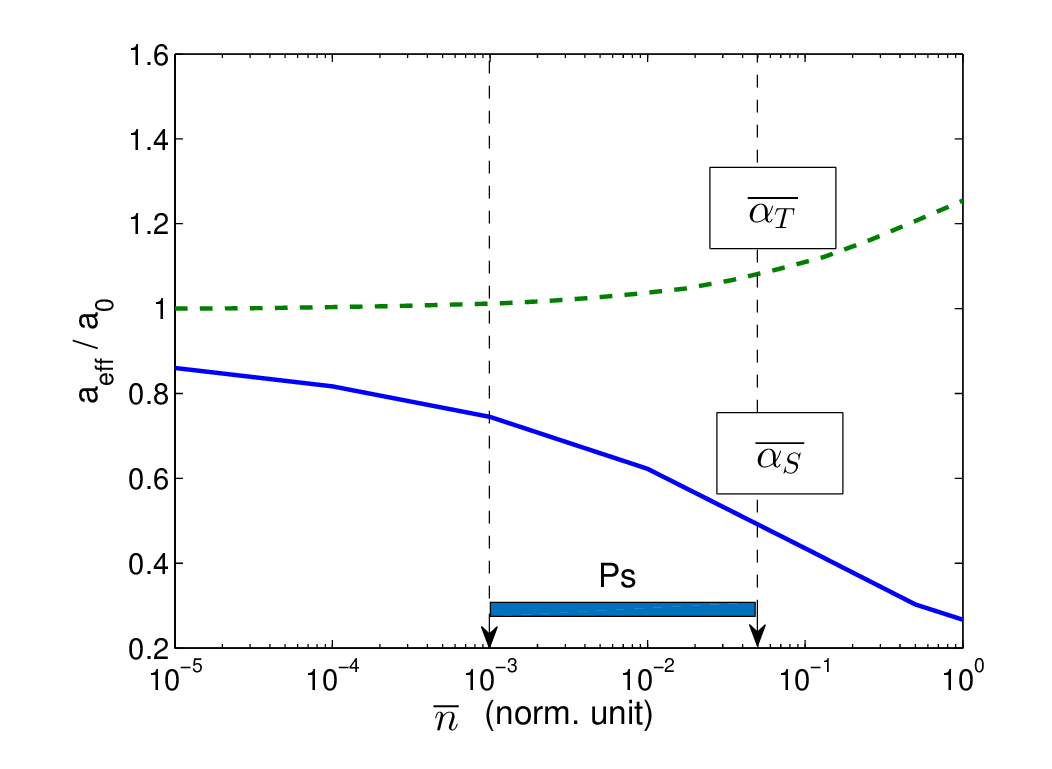}
\caption{\label{fig 3p_sig_Einf} Left panel: Effective scattering length $\za_{\textsf{S} }$ as a function of the energy,  for different  normalized condensate densities $\overline{n}$. Right panel: Mean effective scattering length for NN (green dashed curve) and NC (blue continuous curve).}
\end{center}
\end{figure}
We apply our minimization procedure to the collision operator $\mathcal{W}$ of Eq. \eqref{W}. We note that $\mathcal{W}$ depends linearly on the condensate density $n_c$ and on the bare scattering length $a_0$. In analogy with the previous section, we define the minimization function $\zs$ in Eq. \eqref{zs gen for} by $\zs\equiv \za_{\textrm{S}} $ and we interpret the quantity $a_0\za_{\textrm{S}}$ as the effective scattering length for the NC interaction.  Equation \eqref{zs gen for}  leads to
\begin{align}
\za_{\textsf{S}}(E_1) = \frac{\mathcal{W}^l [\textsf{S},f_{BE}] }{\mathcal{W}^l [1,f_{BE}]}
=\frac{   \int_{-\infty} ^{\infty}\textsf{S} (E_1,|E_2|,E_1+ E_2)  \mathcal{M}''   \dif E_2  }{ \int_{-\infty} ^{\infty} \frac{    \mathcal{M}''  }{\sqrt{1+\left(\frac{\overline{n}}{ E_2}\right)^2}  \sqrt{1+\left(\frac{\overline{n}}{E_1+ E_2}\right)^2 }} \dif E_2  } \; ,\label{za NC}
\end{align}
where he have defined
\begin{align*}
\mathcal{M}''  =       \left(1-e^{-E_1 - E_2}\right)^{-1}   \left\{\begin{array}{ll}
                                            2     \left(e^{E_2}-1\right)^{-1}  &     E_2 > 0         \\
                                                   \left(1-e^{E_2}\right)^{-1}       &    -E_1 <E_2 < 0
                                                       \end{array}\right. \; .
\end{align*}
The right-hand side of Eq. \eqref{za NC} follows from Eq. \eqref{Wpm final form} after some straightforward algebra. The result of the calculation is displayed in Fig. \ref{fig 3p_sig_Einf}. In the left panel we plot $\za_{\textrm{S}} $ as a function of the boson energy for different values of the parameter $\overline{n}$. In contrast to the NN collisions, here the picture of a two-population behavior does not apply. All the curves show a minimum of the interaction for $E \simeq \overline{n}$ and the limit of zero energy is $\za_{\textrm{S}} (0)=0.5$. It is easy to show that the limit of vanishing condensate density gives $\lim_{\overline{n} \rightarrow 0}\za_{\textrm{S}} =1 $  as expected. However,  the convergence is slow and even for small condensate densities (for example $\overline{n} =10^{-5}$) $\za_{\textrm{S}}$ differs significantly from the unity. This fact has a direct consequence on the growth rate of the condensate. The evolution equation of the condensate density is
\begin{align*}
\dpt{n_c}{t}  =-\frac{2}{\hbar}       \int \mathcal{W} [\textsf{S},f]  \dif\mathbf{p}_1 \;.
\end{align*}
Our calculations show that, as a first approximation, we can model the condensation as a hard-sphere collision process with scattering length  $a_{eff} = a_0 \za_{\textrm{S}} < a_0$. The direct consequence is that the growth of the condensate proceeds more slowly than would be the case if the bare interaction $  a_0  $ were used.
{The study of the dynamical evolution of the number of condensate particles and the estimate of the speed at which the condensate is formed is particularly relevant for the positronium system. Our results indicate that when the condensate starts to form, the two-body interaction becomes stronger. This eases the transition of the atoms from the gas to the condensate. This behavior was already observed in Ref. \cite{Mor_13PRA}, where the evolution of a condensate obtained by evaporation of a trapped boson gas was reproduced through a kinetic approach. The results showed that  the increase of the  collision interaction strength leads to a more efficient  thermalization of the atoms, thus speeding up the formation of the condensate.}
In analogy with Eq. \eqref{za eff l}, we can quantify the modification of the scattering length by taking the mean of $\za_{\textrm{S}} $ on the total population of noncondensed bosons. We obtain
\begin{align}
  \overline{\za_{\textsf{S}}}=  \frac{1}{n}\int \za_{\textsf{S}}  f_{BE}(\mathbf{p}) \dif \mathbf{p}  =  \frac{1}{n}\int_{0}^\infty \za_{\textsf{S}} (E) \frac{1}{e^E-1}  \zr(E) \dif E\; .
\end{align}
The result is displayed in the right panel of Fig. \ref{fig 3p_sig_Einf} (blue curve). For sake of comparison, in the same plot we also depict the analogous calculation for $\za_{\textsf{T}}$ (similar to the result of Fig. \ref{fig zs_E} but without distinguishing between high and low energy bosons). Comparing with the $\za_{\textsf{T}}$ curve, we see that, in the case of the NC interaction, the variation of the main effective scattering length is more pronounced. For $\overline{n} = 4\times 10^{-2}$  (ortho-positronium density of $10^{-3}$ nm$^{-3}$) the scattering length decreases by around $50 \%$ for the production of the condensate (NC process), while it increases only by few percent for the collisions in the noncondensed gas (NN process).

\section{Conclusions}

We analyzed the boson-boson scattering process below the condensation temperature. Our study was based on the quasi-particle Bogoliubov theory and the two-body collisions are described by the Boltzmann formalism.
By using a variational approach, we approximated the complex scattering interaction with a hard-sphere collision process and found that a modified scattering length should be used.
Such an effective scattering length quantifies the corrections to the bare scattering interaction in a simple manner.
Our results are general and apply to different species of bosons.
The noncondensed bosons can be classified in two groups. The quasi-free bosons that are essentially unaffected by the presence of the condensate and the low-energy bosons for which the scattering length is modified.
The corrections to the bare scattering length are expressed in terms of a single dimensionless parameter that completely characterizes the condensate. The connection of our theory with the condensation dynamics of the positronium was also discussed.

\appendix
\section{Appendix: }

We state here a simplified form of the Boltzmann collision operators given in  Eqs. \eqref{W}-\eqref{Qren}  under the assumption that the boson distribution function is isotropic on the momentum variable. The explicit form of Eq. \eqref{Qren} has already been derived in other publications \cite{Morandi_14EJPD}. Here, for ease of the reader, we state the final result
%\begin{align*}
% \mathcal{Q} [\textsf{T}_{cn},f]   =&    \frac{g^2 m^34\pi^2}{ \mathcal{E}^{-1}\left(E_1 \right)} \int_0^{\infty}    \int_0^{E_1+E_2}   \left[\left(1+f_1 \right) \left(1+ f_2\right)  f_3  f(E_1+E_2-E_3)  -f_1  f_2 \left(1+f_3\right) \left(1+  f(E_1+E_2-E_3)\right) \right]  \\&  \nn
%             \frac{\textsf{T}_{nn}  \zz (E_1,E_2,E_3)    }{\sqrt{ 1 + \left(\frac{g n_c}{E_2} \right)^2 } \sqrt{ 1 + \left(\frac{g n_c}{E_3} \right)^2 }    \sqrt{ 1 + \left(\frac{g n_c}{E_1+E_2-E_3} \right)^2 }    }    \dif E_2 \dif E_3
%\end{align*}
\begin{align}
 \mathcal{Q}[\textsf{T},f]   = &  \zg \int \textsf{T} \left[\left(1+f_1 \right) \left(1+ f_2\right)  f_3  f_4  -f_1  f_2 \left(1+f_3\right) \left(1+  f_4\right) \right]   \nn  \\
 & \hspace{2cm}\times \zd\left(E_1 + E_2- E_3 - E_4 \right) \zd \left(\mathbf{p}_1 +\mathbf{p}_2-\mathbf{p}_3-\mathbf{p}_4\right) \dif\mathbf{p}_2 \dif\mathbf{p}_3 \dif\mathbf{p}_4  \nn  \\
 =&    \frac{\zg }{ p_1  } \int_0^{\infty}    \left[\left(1+f_1 \right) \left(1+ f_2\right)  f_3  f_4  -f_1  f_2 \left(1+f_3\right) \left(1+  f_4\right) \right] \textsf{T}    \; \zz  \nn \\&
                    \zd\left(E_1+E_2-E_3-E_4\right) \dif E_2 \dif E_3 \dif E_4 \;, \label{Q simp}
\end{align}
where
\begin{align}
\zz (E_1,E_2,E_3,E_4)  = & \left[\min \left( p_1 + p_3 ,p_2 +p_4 \right)-\max \left(\left|p_1 -p_3 \right|,\left|p_2 -p_4 \right|\right)\right] \prod_{i=2,3,4}  \frac{1 }{\sqrt{ 1 + \left(\frac{\overline{n}}{E_i} \right)^2 }}  \;.\label{zz}
\end{align}
We denoted by $p_i= \mathcal{E}^{-1} (E_i)=    \sqrt{\sqrt{E^2 +  (\overline{n})^2}  -   \overline{n}} $, the inverse of the Bogoliubov energy dispersion. The collision kernel is given by
\begin{align}
 \textsf{T} (E_1,E_2,E_3,E_4) &=&  \left( u_1u_2 u_3 u_4 +  v_1u_2 u_3 u_4  + v_1u_2 u_3 v_4+ v_1v_2 v_3 u_4 + u_1 v_2 v_3 v_4   \right)^2\; , \label{T}
\end{align}
where
\begin{eqnarray*}
 u^2_i&=&   \frac{1}{2}+ \frac{E_i}{2\sqrt{E^2_i+(\overline{n})^2}}
\end{eqnarray*}
and $v=\sqrt{1-u^2}$. The derivation of Eq. \eqref{T} can be found in Ref. \cite{Mor_13PRA} or in a equivalent form in Ref. \cite{Imamovic_01}.

We derive the explicit form of the NC Boltzmann kernel. We start from
\begin{align}
 \mathcal{W} [\textsf{S},f]   =&  2\xi \int    \zd\left(E_1 + E_2 - E_3  \right)   \zd \left(\mathbf{p}_1 +\mathbf{p}_2-\mathbf{p}_3 \right)  \left[ \left(1+f_1 \right) \left(1+ f_2\right)  f_3 - f_1  f_2 \left(1+f_3\right)  \right] \textsf{S}  \dif\mathbf{p}_2 \dif\mathbf{p}_3    \nn  \\
%%%
&  +\xi  \int   \zd\left( E_1 - E_2 - E_3 \right) \zd \left(\mathbf{p}_1 -\mathbf{p}_2 -\mathbf{p}_3\right) \left[  \left(1+f_1 \right)  f_2  f_3 - f_1  \left(1+f_2\right) \left(1+f_3\right)  \right]\textsf{S}  \dif\mathbf{p}_2 \dif\mathbf{p}_3  \;.   \nn   %
\end{align}
The collision kernel $ \textsf{S}$ is given by \cite{Mor_13PRA,Griffin_book,Kirkpatrick_85}
\begin{align}
 \textsf{S} (E_1,E_2,E_3) &=   \left( {u}_1 {u}_2 {u}_3 + {v}_1 {v}_2 {v}_3 + {u}_1 {v}_2 {v}_3  + {v}_1  {u}_2 {v}_3 -{u}_1  {v}_2 {u}_3  - {v}_1 {u}_2 {u}_3  \right)^2\;. \label{S}
\end{align}
 We make the substitutions $ \mathbf{p}_2\rightarrow -\mathbf{p}_2$ and we write in a compact form
%\begin{align*}
%\mathcal{W}_{cn}  =& 2 \zg \int     \zd\left(E_1 + E_2 - E_3  \right)    \zd \left(\mathbf{p}_1 +\mathbf{p}_2+\mathbf{p}_3 \right)    \left[ \left(1+f_1 \right) \left(1+ f_2\right)  f_3 - f_1  f_2 \left(1+f_3\right)  \right] \textsf{T}_{cn}  \dif\mathbf{p}_2 \dif\mathbf{p}_3  \\
% %
% &+ \zg \int    \zd\left( E_1 - E_2 - E_3 \right)    \zd \left(\mathbf{p}_1 +\mathbf{p}_2+\mathbf{p}_3 \right)  \left[ \left(1+f_1 \right) f_2  f_3 - f_1 \left(1+  f_2\right) \left(1+f_3\right)  \right] \textsf{T}_{cn}  \dif\mathbf{p}_2 \dif\mathbf{p}_3
%\end{align*}
\begin{align*}
 \mathcal{W} [\textsf{S},f]   =&  \mathcal{W}_+ [\textsf{S},f]  +\mathcal{W}_- [\textsf{S},f] \;.
\end{align*}
We obtain
%\begin{align*}
% \mathcal{Q}^\pm [h] ( E_1)=&\zg \int    \zd\left( E_1 \pm E_2 - E_3 \right)    \zd \left(\mathbf{p}_1 +\mathbf{p}_2+\mathbf{p}_3 \right) \\& \left[ \left(1+f_1 \right) \left(\theta^\pm+ f_2\right)  f_3 - f_1 \left(\theta^\mp +  f_2\right) \left(1+f_3\right)  \right] \textsf{T}_{cn}  \dif\mathbf{p}_2 \dif\mathbf{p}_3
%\end{align*}
\begin{align}
  \mathcal{W}_\pm [\textsf{S},f] ( E_1)=&  \xi  \int \left[ \left(1+f_1 \right) \left(\theta^\pm+ f_2\right)  f_3 - f_1  \left(\theta^\mp +  f_2\right) \left(1+f_3\right)  \right] \textsf{S}\;  p_2^2\;p_3^2\;
   \nn  \\
 &  \times  \zd\left(E_1\pm E_2 - E_3  \right)  \zd \left(\mathbf{p}_1 +\mathbf{p}_2+\mathbf{p}_3 \right)\dif p_2\dif p_3  \dif\widehat{\mathbf{p}}_2 \dif\widehat{\mathbf{p}}_3  \;,   \label{QQ2}
\end{align}
and $\theta^+ \equiv 1$, $\theta^- \equiv 0$. We used $\dif\mathbf{p} = p^2 \dif p \dif \widehat{\mathbf{p}}$ and the hat denotes the unit vector. We consider the angular integration in Eq. \eqref{QQ2}
\begin{align*}
\mathcal{I} =&  \int \zd \left(\mathbf{p}_1 +\mathbf{p}_2+\mathbf{p}_3 \right)\dif\widehat{\mathbf{p}}_2 \dif\widehat{\mathbf{p}}_3 =  \frac{1}{(2\pi)^3}\int  e^{i\left(\mathbf{p}_1 +\mathbf{p}_2+\mathbf{p}_3 \right)\cdot {\bs \zh} } \zh^2 \dif  \zh \dif \widehat{{\bs \zh}} \dif\widehat{\mathbf{p}}_2 \dif\widehat{\mathbf{p}}_3 .
\end{align*}
%%%%%%%%%%%%%%%%%%%%%
Using
\begin{align}
 &  \int   e^{i \mathbf{p}\cdot  {\bs \zh} } \dif \widehat{\mathbf{p}} =  \int   e^{i \mathbf{p}\cdot  {\bs \zh} } \dif \widehat{{\bs \zh}} = 2\pi \int_0^\pi  e^{i p \zh \cos \phi }   \sin \phi \dif \phi  = 4 \pi \frac{\sin p \zh}{p \zh},
 \end{align}
where $\phi$ denotes the angle between $ \mathbf{p}$ and $ {\bs \zh}$, we obtain
\begin{align*}
\mathcal{I} =&   \frac{1}{(2\pi)^3}\int  e^{i\left(\mathbf{p}_1 +\mathbf{p}_2+\mathbf{p}_3 \right)\cdot {\bs \zh} } \zh^2 \dif  \zh \dif \widehat{{\bs \zh}} \dif\widehat{\mathbf{p}}_2 \dif\widehat{\mathbf{p}}_3 = \frac{8}{p_1p_2p_3} \int_0^\infty \frac{\sin (p_1 \zh)\sin (p_2 \zh)\sin (p_3 \zh)  }{\zh}\dif  \zh.
\end{align*}
Using $
 \int_0^\infty  \frac{\sin (a \zh)}{\zh}\dif  \zh =   \frac\pi2 \textrm{sgn} (a)
$  where $\textrm{sgn} $ denotes the sign,
%\begin{align*}
% \sin (p_1 \zh)\sin (p_2 \zh)\sin (p_3 \zh)  =-\frac{1}{4} (\sin [(p_1+p_2+p_3) \zh]+\sin [(p_1-p_2-p_3) \zh]+\sin [(p_2-p_1-p_3) \zh]+\sin [(p_3-p_1-p_2) \zh])
%\end{align*}
%we have
%\begin{align*}
% \int_0^\infty  \frac{\sin (p \zh)}{\zh}\dif  \zh = \frac{1}{2}  \int_{-\infty}^\infty  \frac{\sin (p \zh)}{\zh}\dif  \zh = \frac{1}{2}  \textrm{Im} \int_{-\infty}^\infty  \frac{e^{i p \zh}}{\zh}\dif  \zh =  \frac\pi2 \textrm{sgn} (p)  .
%\end{align*}
we obtain
%\begin{align*}
%\mathcal{I} =& - \frac{ \pi}{p_1p_2p_3} \left[\textrm{sgn}(p_1+p_2+p_3) +\textrm{sgn} (p_1-p_2-p_3)+\textrm{sgn} (p_2-p_1-p_3) +\textrm{sgn} (p_3-p_1-p_2) \right] \\
%=& - \frac{ \pi}{p_1p_2p_3} \left[1 +\textrm{sgn} (p_1-p_2-p_3)+\textrm{sgn} (p_2-p_1-p_3) +\textrm{sgn} (p_3-p_1-p_2) \right]\;.
%\end{align*}
\begin{align*}
\mathcal{I} =& - \frac{ \pi}{p_1p_2p_3} \left[1 +\textrm{sgn} (p_1-p_2-p_3)+\textrm{sgn} (p_2-p_1-p_3) +\textrm{sgn} (p_3-p_1-p_2) \right]\;.
\end{align*}
After simple manipulations, Eq. \eqref{QQ2} becomes
%\begin{align*}
% \mathcal{Q}^\pm [h] ( E_1)=&- \frac{ \zg   \pi}{p_1 }   \int \left[ \left(1+f_1 \right) \left(\theta^\pm +  f_2\right)  f_3 - f_1  \left(\theta^\mp +  f_2\right) \left(1+f_3\right)  \right] \textsf{T}_{cn} p_2 \;p_3 \;
%   \nn  \\
% &  \times   \zd\left(E_1 \pm E_2 - E_3  \right)  \\ & \times \left[1 +\textrm{sgn} (p_1-p_2-p_3)+\textrm{sgn} (p_2-p_1-p_3) +\textrm{sgn} (p_3-p_1-p_2) \right]  \dif p_2\dif p_3 \\
% =&- \frac{ \zg   \pi}{p_1 }   \int  \left[ \left(1+f_1 \right) \left(\theta^\pm +  f_2\right)  f_3 - f_1  \left(\theta^\mp +  f_2\right) \left(1+f_3\right)  \right] \textsf{T}_{cn} \zz_2 \;\zz_3 \;
%   \nn  \\
% &  \times    \zd\left(E_1 \pm E_2 - E_3  \right) \\ & \times \left[1 +\textrm{sgn} (p_1-p_2-p_3)+\textrm{sgn} (p_2-p_1-p_3) +\textrm{sgn} (p_3-p_1-p_2) \right]  \dif E_2\dif E_3 \;,
%\end{align*}
%
\begin{align*}
 \mathcal{W}_+ [\textsf{S},f] ( E_1)=& -\frac{ \xi   \pi}{p_1 }   \int_0^\infty \left[ \left(1+f_1 \right) \left(1+ f_2\right)  f_3 - f_1  f_2 \left(1+f_3\right)  \right] \textsf{S}  \chi_2 \;\chi_3   \\ & \times \left[1 +\textrm{sgn} (p_1-p_2-p_+)+\textrm{sgn} (p_2-p_1-p_+) +\textrm{sgn} (p_+-p_1-p_2) \right]  \dif E_2\;,
\end{align*}
where $p_i=\mathcal{E}^{-1}(E_i)$ (see above), $p_+ = \mathcal{E}^{-1}(E_1 + E_2)$ and  $\chi =\frac{1}{2\dpt{E}{p^2}}=\frac{m}{\sqrt{1+\left(\frac{gn_c}{E}\right)^2 }}$.
%\begin{align*}
% \mathcal{Q}^+ [h] ( E_1)=& -\frac{ \zg   \pi}{p_1 }   \int \left[ \left(1+f_1 \right) \left(1+ f_2\right)  f_3 - f_1  f_2 \left(1+f_3\right)  \right] \textsf{T}_{cn} \zz_2 \;\zz_3   \zd\left(E_1 + E_2 - E_3  \right)  \\ & \times \left[1 +\textrm{sgn} (p_1-p_2-p_3)+\textrm{sgn} (p_2-p_1-p_3) +\textrm{sgn} (p_3-p_1-p_2) \right]  \dif E_2\dif E_3 \\
% =& -\frac{ \zg   \pi}{p_1 }   \int_0^\infty \left[ \left(1+f_1 \right) \left(1+ f_2\right)  f_3 - f_1  f_2 \left(1+f_3\right)  \right] \textsf{T}_{cn} \zz_2 \;\zz_3   \\ & \times \left[1 +\textrm{sgn} (p_1-p_2-p_+)+\textrm{sgn} (p_2-p_1-p_+) +\textrm{sgn} (p_+-p_1-p_2) \right]  \dif E_2\;,
%\end{align*}
By using $\mathcal{E} (p_1 - p_2)\leq \mathcal{E} (p_1)\pm\mathcal{E} (p_2)\leq \mathcal{E} (p_1 + p_2)$ for $p_1\geq p_2$, the previous integral can be easily simplified. We obtain
\begin{align}
 \mathcal{W}_\pm [\textsf{S},f]  ( E_1)=& \frac{ \xi  2\pi m^2 }{p_1 }  \int_0^{\infty}  \int_0^{\infty}\left[ \left(1+f_1 \right) \left(\theta^\pm +  f_2\right)  f_3 - f_1  \left(\theta^\mp +  f_2\right) \left(1+f_3\right)  \right]\nn \\ &\times  \frac{\textsf{S} (E_1,E_2,E_3) (\theta^\pm+1) }{\sqrt{1+\left(\frac{gn_c}{ E_2}\right)^2 }\sqrt{1+\left(\frac{gn_c}{E_3}\right)^2 }}  \zd \left(E_1\pm E_2-E_3\right)   \dif E_2 \dif E_3 \;.\label{Wpm final form}
\end{align}


\begin{thebibliography}{99}



\bibitem{Karshenboim_04}  S. G. Karshenboim, Int. J. Mod. Phys. A \textbf{19}, 3879 (2004).

\bibitem{Guessoum _91} N. Guessoum, R. Ramaty and R. E. Lingenfelter, Astrophys. J. \textbf{378}, 170 (1991).

\bibitem{Cassidy_08} Cassidy and P. Mills, Phys. Rev. Lett. \textbf{100 }, 013401 (2008).

\bibitem{Nagashima_95} Y. Nagashima, M. Kakimoto, T. Hyodo, K. Fujiwara, A. Ichimura, T.  Chang, J. Deng, T. Akahane, T. Chiba, K. Suzuki, B. T. A. McKee, A. T. Stewart, Phys. Rev. A {\bf 52}, 258 (1995).

\bibitem{Puska_94} M. J. Puska, R. M. Nieminen, Rev. Mod. Phys. \textbf{66}, 841 (1994).

\bibitem{Gorgol_12}  M. Gorgol, M. Tydda, A. Kierys, R. Zaleski,  Microporous and Mesoporous Materials \textbf{163}, 276 (2012).

\bibitem{Ferragut_10} R. Ferragut at al.,   Journal of Physics: Conf. Series \textbf{225},  012007  (2010).

\bibitem{Indelicato_14}  Indelicato et al., Hyperfine Interact.  \textbf{228}, 141 (2014).


\bibitem{Cassidy_07b} D. B. Cassidy and A. P. Mills,   Nature \textbf{449}, 7159: 195 (2007).

\bibitem{Fetter_02} A. L. Fetter, Journal of Low Temp. Phys. {\bf 129}, 263 (2002).

\bibitem{Platzman_94} P. M. Platzman and A. P. Mills,  Phys. Rev. B {\bf 49}, 454 (1994).

\bibitem{Mills_15} A. P. Mills, J. of Phys.: Conf. Series \textbf{505},  012039  (2014).

\bibitem{Kasprzak_06} J. Kasprzak et al., Nature \textbf{443}, 409 (2006).

\bibitem{Cassidy_10} D. B. Cassidy, V. E. Meligne and A. P. Mills, Phys. Rev. Lett. \textbf{104}, 173401 (2010).

\bibitem{Cassidy_10a} D. B. Cassidy, P. Crivelli, T. H. Hisakado, L. Liszkay, V. E. Meligne, P. Perez, H. W. K. Tom and A. P. Mills Jr, Phys. Rev. A \textbf{81}, 012715 (2010).
    
\bibitem{Morandi_14EJPD}  O. Morandi, P.-A. Hervieux, and G. Manfredi,  Eur. Phys. J. D \textbf{68}, 84 (2014).


\bibitem{Nagashima_98} Y. Nagashima, T Hyodoy, K Fujiwarayx and A Ichimuraz, J. Phys. B: At. Mol. Opt. Phys. {\bf 31}, 329 (1998).

\bibitem{He_07} C. He, T. Ohdaira, N. Oshima, M. Muramatsu, A. Kinomura, R. Suzuki, T. Oka and Y. Kobayashi, Phys. Rev. B \textbf{75}, 195404 (2007).

\bibitem{Zaleski_13}  R. Zaleski, J. of Phys.: Conf. Series \textbf{443}, 012062 (2013)

\bibitem{Mor_14PRA} O. Morandi, P.-A. Hervieux, and G. Manfredi,  Phys. Rev. A \textbf{89}, 033609 (2014).

\bibitem{Mor_14JPB} O. Morandi, P.-A. Hervieux, and G. Manfredi, J. Phys. B \textbf{47}, 155202 (2014).

\bibitem{Mariazzi_10} S. Mariazzi, P. Bettotti and R. S. Brusa, Phys. Rev. Lett. {\bf104}, 243401 (2010).

\bibitem{Ferragut_13} R. Ferragut, S. Aghion,G. Tosi, G. Consolati,M. Longhi,A. Galarneau. F. Di Renzo, J. Phys. Chem. C, \textbf{117}, 26703 (2013).

\bibitem{Andersen_87}  S. L. Andersen, D. B. Cassidy, J. Chevallier, B. S. Cooper, A. Deller, T. E. Wall and U. I. Uggerhoj, J. Phys. B: At. Mol. Opt. Phys. \textbf{48},  204003 (2015).

%
\bibitem{Oka_14} T. Oka, Y. Sano, Y. Kino, T. Sekine, Eur. Phys. J. D \textbf{68}, 156 (2014).

\bibitem{Murtagh_09} D. J. Murtagth, D. A. Cooke and G. Laricchia,  Phys. Rev. Lett. \textbf{102}, 133202 (2009).


\bibitem{Shibuya_13} K. Shibuya, T. Nakayama, H. Saito, T. Hyodo,  Phys. Rev. A, \textbf{88}, 012511 (2013).

\bibitem{Mor_15JP} O. Morandi  and P.-A. Hervieux, J. Phys.: Conf. Series \textbf{618}, 012011 (2015).


\bibitem{Takada_00} S. Takada, T. Iwata, K. Kawashima, H. Saito, Y. Nagashima, T. Hyodo, Radiat. Phys. Chem. (UK), \textbf{58}, 781  (2000).

\bibitem{Chang_87} T. B. Chang, M. Xu and X. Zeng, Phys. Lett. A \textbf{126}, 189 (1987).

\bibitem{Griffin_book} A. Griffin, T. Nikuni, E. Zaremba, {\it Bose condensed gases at finite temperatures}, Cambridge University Press (2009).

\bibitem{Bogoliubov_47} N. N. Bogoliubov, J. Phys. USSR \textbf{11}, 23,  189 (1947).

\bibitem{Baliev_58}  S. T. Beliaev, Sov. Phys. JETP \textbf{7}, 289 (1958).

\bibitem{Popov_87} V. N. Popov, Sov. Phys. JETP \textbf{20},  1185 (1965). 

\bibitem{Kirkpatrick_85} T. R. Kirkpatrick and J. R. Dorfman, J. Low Temp. Phys. {\bf 58}, (3-4), 301 (1985).

\bibitem{Jackson_02} B. Jackson and E. Zaremba,   Phys. Rev. A. {\bf 66}, 033606 (2002).

\bibitem{Mor_13PRA} O. Morandi, P.-A. Hervieux, and G. Manfredi, Phys. Rev. A \textbf{88}, 23618 (2013).

\bibitem{Imamovic_01} M. Imamovic-Tomasovic and A. Griffin, J. of Low Temp. Phys. {\bf 122}, 617 (2001).

\bibitem{Banyai_00} L. B\'anyai, P. Gartner, O. M. Schmitt, and H. Haug, Phys. Rev. B \textbf{61}, 8823 (2000).

\bibitem{Ivanov_02} I. A. Ivanov, J.  Mitroy and  K. Varga, Phys. Rev. A. {\bf 65}, 022704 (2002).

\bibitem{Pethick_book} C. J. Pethick, H. Smith, {\it Bose-Eintein Condensation in Dilute Gases}, Cambridge University Press (2004).





\end{thebibliography}
\end{document}